\documentclass[aip,jcp,superscriptaddress,amsmath,amssymb,twocolumn,reprint]{revtex4-1}

\usepackage[version=3]{mhchem} 
\usepackage{graphicx}
\usepackage{amsmath}
\usepackage{amssymb}
\usepackage{nicefrac}
\usepackage{booktabs}
\usepackage{array}
\usepackage{ulem}

\def\rv{{\bf r}}
\def\fv{{\bf f}}
\def\dd{\mathrm{d}}

\def\beq{\begin{equation}}
\def\eeq{\end{equation}}

\newcommand{\br}{\mathbf{r}}
\newcommand{\bu}{\mathbf{u}}

\usepackage{color}
\usepackage{pstricks}
\usepackage{ifthen}
\newrgbcolor{stefancolor}{.1 .1 .7}
\newrgbcolor{paolacolor}{.7 .1 .1}
\newrgbcolor{augucolor}{.5 .1 .7} 
\usepackage{csquotes}

\newboolean{DEBUG}
\setboolean{DEBUG}{true}

\ifthenelse {\boolean{DEBUG}}
{}

\ifthenelse {\boolean{DEBUG}}
{}

\ifthenelse {\boolean{DEBUG}}
{}

\ifthenelse {\boolean{DEBUG}}
{}

\ifthenelse {\boolean{DEBUG}}
{}

\ifthenelse {\boolean{DEBUG}}
{}

\begin{document}    


\author{Stefan Vuckovic}
\email{stefanvuckovic1@gmail.com}
\affiliation
{Institute for Microelectronics and Microsystems (CNR-IMM), 73100 Lecce Italy}
\affiliation{Department of Chemistry \& Pharmaceutical Sciences and Amsterdam Institute of Molecular and Life Sciences (AIMMS), Faculty of Science, Vrije Universiteit, De Boelelaan 1083, 1081HV Amsterdam, The Netherlands}

\author{Augusto Gerolin}
\email{agerolin@uottawa.ca}
\affiliation
{Department of Chemistry and Biomolecular Sciences, University of Ottawa}
\affiliation
{Department of Mathematics and Statistics, University of Ottawa}

\author{Timothy J. Daas}
\email{t.j.daas@vu.nl}
\affiliation
{Department of Chemistry \& Pharmaceutical Sciences and Amsterdam Institute of Molecular and Life Sciences (AIMMS), Faculty of Science, Vrije Universiteit, De Boelelaan 1083, 1081HV Amsterdam, The Netherlands}

\author{Hilke Bahmann}
\email{bahmann@uni-wuppertal.de}
\affiliation{Physical and Theoretical Chemistry, University of Wuppertal, Gausstr. 20 42119 Wuppertal}

\author{Gero Friesecke}
\email{gf@tum.de}
\affiliation
{Department of Mathematics, Technische Universit\"at M\"unchen}

\author{Paola Gori-Giorgi}
\email{p.gorigiorgi@vu.nl}
\affiliation
{Department of Chemistry \& Pharmaceutical Sciences and Amsterdam Institute of Molecular and Life Sciences (AIMMS), Faculty of Science, Vrije Universiteit, De Boelelaan 1083, 1081HV Amsterdam, The Netherlands}

\title{Density Functionals based on the mathematical structure of the strong-interaction limit of DFT}

\begin{abstract}
    While in principle exact, Kohn-Sham density functional theory -- the workhorse of computational chemistry -- must rely on approximations for the exchange-correlation functional. Despite staggering successes, present-day approximations still struggle when the effects of electron-electron correlation play a prominent role. The limit in which the electronic Coulomb repulsion completely dominates the exchange-correlation functional offers a well-defined mathematical framework that provides insight for new approximations able to deal with strong correlation. In particular, the mathematical structure of this limit, which is now well-established thanks to its reformulation as an optimal transport problem, points to the use of very different ingredients (or features) with respect to the traditional ones used in present approximations. We focus on strategies to use these new ingredients to build approximations for computational chemistry and highlight future promising directions.
\end{abstract}

\maketitle

\section{Introduction}
\label{sec_intro}

Owing to its high accuracy-to-cost ratio, Kohn-Sham density functional theory (KS DFT) is presently the primary
building block of successes of quantum chemistry in disciplines that stretch from biochemistry to materials
science. \cite{Bur-JCP-12,PGB15,MH17,GHBE17,VT20,Sim2022} 
DFT calculations consume a significant fraction of the world’s supercomputing power \cite{Sherrill2020}
and tens of thousands of scientific papers report DFT calculations with the number ever growing. \cite{PGB15} 
KS DFT is in principle exact, but in practice it requires approximations to one piece of the total energy, the so-called exchange-correlation (XC) functional, which encodes the quantum,  fermionic and Coulombic nature of electrons.

The construction of modern XC approximations draws from different approaches. 
Some of them are
based on 
forms fulfilling
 some known exact constraints, \cite{Bur-JCP-12,SunRuzPer-PRL-15} some have been fitted to large
databases,\cite{MH17,VT20} and the most recent XC approximations are machine learned. \cite{Kalita2021,fn_dm21,Nagai2020}
Regardless of these differences in
their design, nearly all current DFT approximations are constructed
from the same ingredients (or features)
that form the ``Jacob’s ladder''.\cite{PerSch-AIP-01,Ham-SCI-17} 

Despite the progress,\cite{fn_dm21} state-of-the-art XC approximations have been greatly successful mainly in describing only weak and moderate electronic correlations. \cite{MH17,GHBE17} The inability of state-of-the-art DFT to capture strong correlations hampers its reliability and predictive power. \cite{Bur-JCP-12,CohMorYan-CR-12,PGB15,Sim2022}
Over the last two decades,  the strongly interacting limit of DFT (SIL)  \cite{Sei-PRA-99,SeiGorSav-PRA-07,GorVigSei-JCTC-09,ButDepGor-PRA-12,FriGerGor-arxiv-22} has been explored and 
a rigorous theory has been established. 
This theory reveals mathematical objects that are very different from the ingredients that are used for building standard XC approximations (semilocal quantities and KS orbitals forming the Jacob's ladder). By offering building blocks for XC functionals
tailored to describe strong correlations, the SIL has a potential to solve the long-standing problem of DFT simulations of strong electronic correlations. 

Here we give a summary of the development of the SIL in different contexts:
the development of the theory itself, its practical realization, and the development of approximations drawing from it.
We discuss paths for using this limit in different ways to solve the problem of strong correlations within DFT and discuss how it has enabled the construction of a range of quantities that can guide the further development of DFT. We also give an overview of how the SIL has motivated the development of methods that go beyond DFT, such as wavefunction methods delivering highly accurate noncovalent interactions.\cite{DaaFabDelGorVuc-JPCL-21}

\section{Theory}
\label{sec_theory}

\subsection{Exchange-correlation functional in DFT}
\label{sec_xc}

Using the Levy-Lieb (LL) constrained-search formalism\cite{Lev-PNAS-79,Lie-IJQC-83} the ground state energy and density of a many-electron system with an external potential $v$ can be obtained as
\begin{equation}\label{eq:gs}
E_{\rm GS}[v] = \min_\rho \left\lbrace F[\rho]+\int v(\rv) \rho (\rv) \mathrm{d}\rv \right\rbrace,
\end{equation}
where $F[\rho]$ is the $\lambda=1$ (physical) value of the generalized universal LL functional for arbitrary coupling constant $\lambda$,
\begin{equation}\label{eq.LLalpha}
F^{\lambda}[\rho]= \min_{\Psi\mapsto\rho} \langle \Psi|\hat{T} +\lambda\,\hat{V}_{ee} | \Psi \rangle.
\end{equation}
In the Kohn-Sham formalism, the functional $F[\rho]=F^{\lambda=1}[\rho]$ is partitioned as
\begin{equation}\label{eq:funi_part}
F[\rho]= T_s[\rho]+U[\rho]+E_{xc}[\rho],
\end{equation}
where $T_s[\rho]=F^{\lambda=0}[\rho]$ is the KS non--interacting kinetic energy, $U[\rho]$ is the Hartree (mean-field) energy, and $E_{xc}[\rho]$ is the exchange--correlation energy.
The adiabatic connection (AC) formula for the XC functional reads\cite{LanPer-SSC-75,GunLun-PRB-76}
\begin{equation}
E_{xc}[\rho]= \int_0^1 W_\lambda[\rho]\mathrm{d}\lambda,
		\label{eq:ac_xc}
\end{equation}
where $W_\lambda[\rho]$ is the global AC integrand:
\begin{equation}
W_\lambda[\rho]= \langle \Psi_{\lambda}[\rho]  | \hat{V}_{ee}| \Psi_{\lambda}[\rho]\rangle - U[\rho],
		\label{eq:w1}
\end{equation}
and $\Psi_{\lambda}[\rho]$ is the minimizing wavefunction in Eq.~\eqref{eq.LLalpha}.
We can also write $W_\lambda[\rho]$ as
\beq \label{eq:wl}
W_\lambda[\rho]=\int w_\lambda(\rv) \rho(\rv) \mathrm{d}\rv,
\eeq
where $ w_\lambda(\rv) $  is a $\lambda$-dependent XC energy density that is not uniquely defined. 
In the present work, we adhere to the  definition 
in terms of the electrostatic potential of the XC hole ({\it conventional DFT gauge}),\cite{BurCruLam-JCP-98,VucIroSavTeaGor-JCTC-16,VucGor-JPCL-17,VucLevGor-JCP-17}
\beq \label{eq:wxc_def}
    w_\lambda(\rv) = \frac{1}{2} \int_0^\infty \frac{h_{\rm xc}^\lambda(\rv,u)}{u}4 \pi u^2 \mathrm{d}u, 
\eeq
where $h_{\rm xc}^\lambda(\rv,u)$ is the spherical average (over directions of ${\bf u}=\rv'-\rv$) of the XC hole around a given position $\rv$. The XC hole, in turn, is determined by the pair density associated to the wavefunction $\Psi_\lambda[\rho]$.

\section{Strongly interacting limit of DFT}
\label{sec_sil}
In this section, we briefly review the physical ideas behind the strongly interacting limit of DFT. 
For a mathematically more rigorous and comprehensive overview, we recommend ref.~\onlinecite{FriGerGor-arxiv-22}.

The strongly interacting limit of DFT corresponds to the situation in which the electron-electron repulsion dominates in $F^{\lambda}[\rho]$ of Eq.~\eqref{eq.LLalpha}, yielding\cite{Sei-PRA-99, SeiGorSav-PRA-07}
\beq \label{eq.asy2}
   \lim_{\lambda\to\infty} \tfrac{1}{\lambda}\, F^\lambda[\rho] = V_{ee}^{\rm SCE}[\rho],
\eeq 
where $V_{ee}^{\rm SCE}[\rho]$
is the {\it strictly correlated electrons (SCE) functional} defined by the minimization of the electronic repulsion over wavefunctions $\Psi$ with density $\rho(\rv)$:
\beq \label{eq.VSCE}
   V_{ee}^{\rm SCE}[\rho] = \min_{\Psi\mapsto\rho}
   \langle\Psi| \hat{V}_{ee}|\Psi\rangle.
\eeq
The limit in eq. \eqref{eq.asy2} has been established rigorously \cite{CotFriKlu-CPAM-13,CotFriKlu-ARMA-18, Lew-CRM-18} with convergence of the `energies', i.e., the value of the functional $F^\lambda[\rho]$ divided by $\lambda$ tends to $V_{ee}^{\rm SCE}[\rho]$, and qualitative convergence of the wave-functions squared, i.e., for any $(\Psi_{\lambda})_{\lambda>0}$  minimizing \eqref{eq.LLalpha} the $N$-body position density $P_\lambda^N=\sum_{\rm spins}|\Psi_\lambda|^2$ converges to a limiting $N$-body  distribution $P^N_\infty[\rho]$. The latter minimizes the following alternative definition of the SCE functional 
\beq
\label{eq:infmin}
   V_{ee}^{\rm SCE}[\rho]
   = \operatorname*{min}_{P^N \mapsto \rho} \int \sum_{i,j\, j>i}^N\frac{1}{r_{ij}} \,P^N(\rv_1,...,\rv_N)\dd \rv_1...\dd \rv_N.
\eeq
Here the minimum is over all symmetric $N$-body probability distributions with one-body density $\rho$. Interestingly, unlike its absolute value squared, the wavefunction $\Psi_\lambda$ itself does not converge to any meaningful limit.
Since $P^N_{\infty}[\rho]$ minimizes only the electronic repulsion, one can think of it as a natural analogue of the Kohn-Sham noninteracting wavefunction $\Psi_0[\rho]$, which minimizes the kinetic energy functional only. 

\vspace{2mm}
\noindent
\textbf{Links to the XC functional:}
The functional 
$V_{ee}^{\rm SCE}[\rho]$ also corresponds to a well defined limit of the XC functional. 
In fact, the $\lambda \to \infty$ limit of the AC integrand of Eq.~\eqref{eq:w1} is equal to
\begin{equation}
W_\infty[\rho]=  V_{ee}^{\rm SCE}[\rho] - U[\rho].
		\label{eq:winfty}
\end{equation}
Moreover, there is a well-known relation\cite{LevPer-PRA-85,LevPer-PRB-93} between scaling the coupling strength $\lambda$ and performing uniform coordinate scaling on the density, $\rho_\gamma (\rv) = \gamma ^3 \rho (\gamma\, \rv) $ (with $\gamma >0$), which implies that the exact XC functional tends to $W_\infty[\rho]$ in the low-density ($\gamma\to 0$) limit. The SCE limit is thus complementary to exchange, which yields the high-density limit ($\gamma\to\infty$) of $E_{xc}[\rho]$,
\begin{equation}
\label{eq:ld-hd}
\lim_{\gamma\to \infty}\frac{E_{xc}[\rho_\gamma]}{\gamma} = E_x[\rho],\qquad  \lim_{\gamma\to 0}\frac{E_{xc}[\rho_\gamma]}{\gamma} = W_\infty[\rho].
\end{equation}



\vspace{2mm}
\noindent
\textbf{The SCE state:} As a candidate for the
 wave-function squared $P^N_\infty[\rho]$, Seidl and co-workers\cite{Sei-PRA-99,SeiGorSav-PRA-07} proposed to restrict the minimization in \eqref{eq:infmin} over singular distributions having the form: 
\begin{equation}\label{eq.psiSCE} 
\begin{split}
  &P^N_{\rm SCE} = \dfrac{1}{N!}\sum_{\mathcal{P}}\int d\mathbf{s}\dfrac{\rho(\mathbf{s})}{N}\delta(\rv_1-\fv_{\mathcal{P}(1)}(\mathbf{s}))\times  \\ 
&\times\delta(\rv_2-\fv_{\mathcal{P}(2)}(\mathbf{s}))\times\dots\times\delta(\rv_N-\fv_{\mathcal{P}(N)}(\mathbf{s})), 
\end{split}
\end{equation}
where $\fv_1,..,\fv_N$ are the so-called {\it co-motion functions}, with $\fv_1(\rv)=\rv$, $\mathcal{P}$ is permutation of $\{1,...N\}$, and $\delta\big(\rv-\fv_i(\rv)\big)$ denotes the delta function of $\rv$ (alias Dirac measure) centered at  $\fv_i(\rv)=\fv_i([\rho];\rv)$. The singular densities \eqref{eq.psiSCE} are concentrated on the d-dimensional set  $\Omega_0\subset \mathbb{R}^{dN}$,
\begin{equation} \label{eq:Omega0}
     \Omega_0 = \{ \rv_2 = f_2(\rv),...,\rv_N=f_{N}(\rv)\},
\end{equation}
and its permutations.
Intuitively speaking, such a $N$-body density describes a state in which the position of one of the electrons, say $\rv\in\mathbb{R}^d$, can be freely chosen according to the density $\rho$, but this then uniquely fixes the position of all the other electrons through the  {\it co-motion maps} $\fv_i(\rv)$, that is, $\rv_2=f_2(\rv), \dots,\rv_N=f_N(\rv)$. Thus states of form \eqref{eq.psiSCE} are  called {\it strictly correlated states}, or SCE states for short.
In other words, if a reference electron is at $\rv$, the other electrons in the SCE state can be found nowhere else, but at the $\fv_i(\rv)$ positions. Besides yielding minimal electronic repulsion, the co-motion functions need to satisfy group properties,\cite{Sei-PRA-99,SeiGorSav-PRA-07,ColDepDim-CJM-15} accounting for the indistinguishability of electrons, and the pushforward conditions for the density constraint.\cite{SeiGorSav-PRA-07,ColDepDim-CJM-15}  

\begin{figure}
\includegraphics[width=\linewidth]{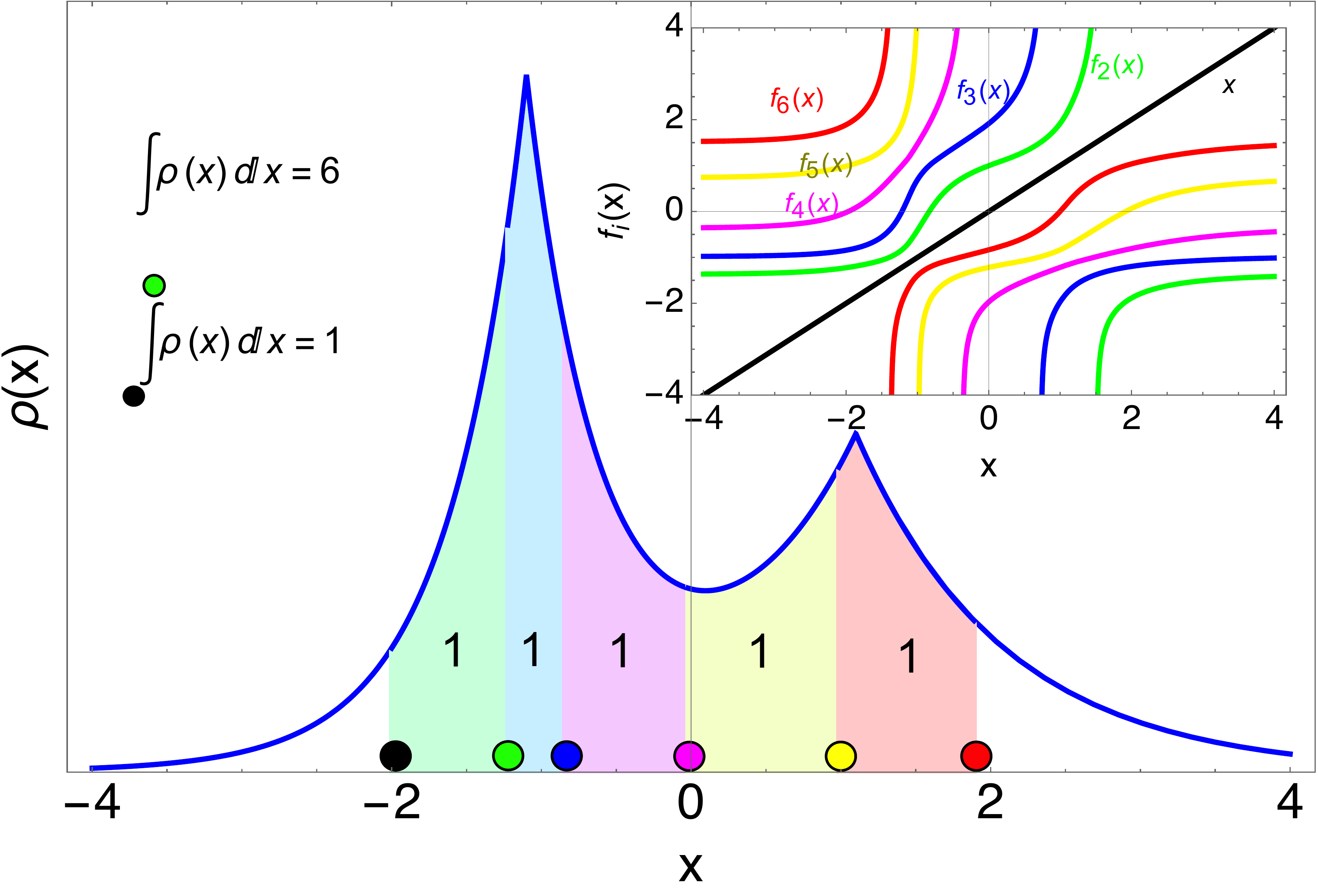}
\caption{A pedagogical example: a sample of strictly correlated positions for a 1D density
integrating to 6 electrons.  
A reference electron is placed at $x=-2$ (black point) 
and the SCE position of the other electrons is determined by 
$f_i(x)$
and are represented by other colors. 
Notice that the shaded area 
between two adjacent SCE positions 
integrates to $1$ and this is what defines $f_i(x)$ for 1D densities. 
Inset is showing the co-motion functions for the given density.}
\label{fig_cmf}
\end{figure}

Constructing the co-motion functions is not simple, except in some special cases such as one-dimensional and spherically-symmetric systems.\cite{Sei-PRA-99,SeiGorSav-PRA-07} In those cases, the co-motion maps are obtained from {\it constrained integrals of the density}.
This is illustrated in Fig.~\ref{fig_cmf}, which shows a simple one-dimensional example of the optimal solution for \eqref{eq.VSCE}, which has the form \eqref{eq.psiSCE}, with strictly correlated positions separated by ``chunks'' of density that integrate to integers.  We should stress that this solution has been rigorously proven to be exact for one-dimensional systems,\cite{ColDepDim-CJM-15} which means that in this case the exact XC functional in the low-density limit is entirely determined by these constrained integrals rather than by any of the traditional Jacob's ladder ingredients.

For finite systems the SCE potential is defined as the functional derivative of the SCE functional $V_{ee}^{\rm SCE}[\rho]$ with respect to the density,
$v_{\rm SCE}(\rv)=\delta V_{ee}^{\rm SCE}[\rho]/\delta \rho(\rv)$, with the convention that $v_{\rm SCE}(\rv)$ tends to zero as $|\rv|\to\infty$.  
Given an SCE state of Eq.~\eqref{eq.psiSCE}, the SCE functional and potential
 can be simply written in terms of the co-motion functions:\cite{MirSeiGor-JCTC-12,MalGor-PRL-12,MalMirCreReiGor-PRB-13}
\begin{align}
	V_{ee}^{\rm SCE}[\rho] & = \int \frac{\rho(\rv)}{2}\sum_{i=2}^{N}\frac{1}{\left | \rv-\fv_i(\rv) \right|}\mathrm{d}\rv 
\label{eq:vee_comotion} \\
\nabla v_{\rm SCE}(\rv) & =-\sum_{i=2}^N \frac{\rv-\fv_i(\rv)}{|\rv-\fv_i(\rv)|^3}. \label{eq:forcev}
	\end{align}
Equation~\eqref{eq:forcev} has a simple physical interpretation:  $v_{\rm SCE}(\rv)$ is the one-body potential that corresponds to the net force exerted on an electron at position $\rv$ by the other $N-1$ electrons.

Is actually the SCE state of  Eq.~\eqref{eq.psiSCE} always the \textit{true} minimizer of Eq.~\eqref{eq:infmin}? It has been proven that this is true when $N=2$ in any dimension $d\geq 1$\cite{CotFriKlu-CPAM-13,ButDepGor-PRA-12} and when $d=1$ for any number of electrons.\cite{ColDepDim-CJM-15} In general, the SCE state of  Eq.~\eqref{eq.psiSCE} is not guaranteed to yield the absolute minimum for the electronic repulsion for a given arbitrary density $\rho(\rv)$.\cite{ColStr-M3AS-15,SeiDiMGerNenGieGor-arxiv-17}
Nevertheless,
numerical evidence suggests that the energetic difference between
$V_{\rm ee}^{\rm SCE} [\rho]$ from the SCE state (Eq.~\eqref{eq:vee_comotion}) 
and the true minimum of Eq.~\eqref{eq:infmin} is very small.\cite{SeiDiMGerNenGieGor-arxiv-17}

\vspace{2mm}
\noindent
\textbf{Other $V_{\rm ee}^{\rm SCE} [\rho]$ formulations.} In addition to  the co-motion functions formulation (Eq.~\ref{eq:vee_comotion}), there are other equivalent formulations for $V_{ee}^{\rm SCE}[\rho]$ arising from mass transporation theory. The link between the SCE functional and mass transportation (or optimal transport) theory was found, independently, by Buttazzo {\it et al.}\cite{ButDepGor-PRA-12} and by Cotar {\it et al.}\cite{CotFriKlu-CPAM-13}.
From the optimal transport viewpoint, the SCE functional defines a multimarginal problem, in which all the marginals are the same, so that the SCE mass-transportation problem corresponds to a reorganization of the ``mass pieces'' within the same density.  From optimal transport theory, the dual Kantorovich formulation 
for $V_{ee}^{\rm SCE}[\rho]$ can be also deduced,\cite{ButDepGor-PRA-12} defining the Kantorovich potential $u(\rv)$,
\begin{align}
    &	V_{ee}^{\rm SCE}[\rho]=	\label{eq:vee_kant} \\ \nonumber
    &		 \max_{u}\left \{  \int u(\rv)\rho (\rv)\dd \rv:\sum_{i=1}^{N}u(\rv_i)\leqslant \sum_{i=1}^{N}\sum_{j>i}^{N}\frac{1}{ |  \rv_{i}-\rv_{j} |}\right \},
\end{align}
 and can be reformulated \cite{MenLin-PRB-13} by a nested optimization akin to the Legendre-Fenchel transform of Eq.~\eqref{eq:infmin}: 
	\begin{align}
	\label{eq:maxmin}
	V_{ee}^{\rm SCE}[\rho ]=\max_{u}\left \{  \int u(\rv)\rho(\rv)\dd\rv+g[u]  \right \},
	\end{align}
with $g[u]$ the minimum 
of the classical potential energy,
\begin{equation}\label{eq:epotmin}
    E^{\rm pot}_{{\rm SCE}}(\rv_1,\dots,\rv_N)=\sum_{i<j}\frac{1}{ |  \rv_{i}-\rv_{j} |}-\sum_{i=1}^N u(\rv_i),
\end{equation}
over $\rv_1,...,\rv_N$.
One can show that the potentials  $u(\rv)$ and $v_{\rm SCE}(\rv)$ differ only by a well-defined constant whose physical meaning has been explored in Refs.~\onlinecite{VucWagMirGor-JCTC-15} and~\onlinecite{VucLevGor-JCP-17}.

\vspace{2mm}

\noindent
\textbf{Next leading term:} More information about the exact LL functional at low density can be gained by studying the next leading term in Eq.~\eqref{eq.asy2}. Under the assumption that the minimizer in \eqref{eq.VSCE} is of the SCE type \eqref{eq.psiSCE}, the classical potential energy \eqref{eq:epotmin} is minimum on the manifold $\Omega_0$ parametrised by the co-motion functions. 
The conjecture is then that the next leading term is given by zero-point oscillations in the directions perpendicular to the SCE manifold\cite{GorVigSei-JCTC-09}
\begin{align} \label{eq:nexttermtot}
F^\lambda[\rho] & \;\underset{\lambda \to \infty}{\sim}  \; \lambda V_{ee}^{\rm SCE}[\rho]+\sqrt{\lambda}\,F^{\rm ZPE}[\rho],
\end{align}\vspace{-6mm}
\begin{equation}
\text{where   }\hspace{2mm}     F^{\rm ZPE}[\rho]=\frac{1}{2}\int_{\mathbb{R}^d} \frac{\rho(\rv)}{N}{\rm Tr}\left(\sqrt{\mathbb{H}(\rv)}\right)\dd\rv,  \label{eq:ZPOfunc}
\end{equation}
and $\mathbb{H}(\rv)$ is the hessian matrix evaluated on $\Omega_0$ for fixed $\rv$. The intuition that this next term should be given by zero-point oscillations around the manifold parametrized by the co-motion functions appeared for the first time in Seidl's seminal work, \cite{Sei-PRA-99} and calculations for small atoms (He to Ne) are reported in Ref.~\onlinecite{GorVigSei-JCTC-09}. A rigorous proof in the one-dimensional case for any $N$ has been provided recently.\cite{ColDMaStra-arxiv-21}

\vspace{2mm}
\noindent
\textbf{The spin state:} Besides the expansion of eq~\eqref{eq:nexttermtot} in terms of powers of $\lambda$, which is semiclassical in nature, it is conjectured\cite{GorVigSei-JCTC-09,GorSeiVig-PRL-09} that the effect of the spin state will enter at large-$\lambda$ through orders $e^{-\sqrt{\lambda}}$, which corresponds to the overlap of gaussians centered in different co-motion functions. This conjecture has been confirmed numerically for $N=2$ electrons in 1D.\cite{GroKooGieSeiCohMorGor-JCTC-17}
\vspace{2mm}

\noindent
\textbf{Numerical realization of the SCE functional:} The SCE functional cannot at the moment be accurately and efficiently computed for general 3D densities and large $N$. But accurate numerical methods are available for small $N$ or special situations,  
and novel methods aimed at large $N$ are under development. In particular, the very recent genetic column generation method  \cite{FriSchVoe-21} appears in test examples to scale favourably with system size.

In Table~\ref{tab_sce} we give an overview of the proposed algorithms for computing the SCE functional and potential
and refer to the book chapter\cite{FriGerGor-arxiv-22} for a more detailed review. From Table~\ref{tab_sce}, we can see that for general 3D densities, numerical solutions were reported only for up to 10 electrons (the first method being limited to radial densities, whereas for the last method 3D tests are not yet available). This  indicates again the level of complexity and ultra non-locality of the SCE functional. 

In fact, in the worst case scenario, the computational complexity of simple algorithms scales exponentially with the number $N$ of electrons\cite{linHoCutJor-arXiv-19} and computing the SCE functional may be NP-hard. \cite{AltBoi-arXiv-20,AltBoi-Dis-21, FriSchVoe-21} In the discrete setting, where the single particle density $\rho$ is supported on $\ell$ points, eq. \eqref{eq:infmin} is equivalent to a linear programming problem with $\ell N$ constraints and $\ell^N$ variables. 

Despite these limitations in solving the SCE problem exactly, rather accurate approximations, retaining some of the SCE non-locality, have been recently proposed and they will be detailed in the next section.

\begin{table*}
\centering
\caption{An overview of proposed SCE algorithms. The third column shows $N_{\rm max}$, which indicates up to how many electrons a given algorithm was applied to. This is not meant as a direct comparison of methods as the reported results differ in the fineness of discretization and the accuracy achieved. 
Also, 
only the second, fourth and last algorithms are free of additional approximations beyond discretization. 
} \label{tab_sce}
    \begin{tabular}{|l|l|l|}
        \hline
        Algorithm                                                    & References                                                                                & $N_{\rm max}$ \\ \hline
         SGS approach based on co-motion functions (radial densities only) &  \onlinecite{Vuc-Thesis-17}, see also \onlinecite{SeiGorSav-PRA-07,SeiVucGor-MP-16} & 100           \\ 
         Linear programming applied to the N-body formulation \eqref{eq:infmin}                     & \onlinecite{CheFriMen-JCTC-14}              & 2             \\
         Multi-marginal Sinkhorn algorithm                            & \onlinecite{BenCarCutNenPey-SIAM-15,BenCarNen-SMCISE-16,DMaGerNen-TOOAS-17,DMaGer-JSC-20,GerGroGor-JCTC-20}                 & 5             \\ 
        Algorithms based on the Kantorovich  formulation \eqref{eq:maxmin}           & \onlinecite{MenLin-PRB-13,VucWagMirGor-JCTC-15}                                           & 6             \\ 
        Algorithm based on representability constraints for the pair density                       & \onlinecite{KhoYin-SIAMJSC-19}                & 10             \\
        Langevin dynamics with moment constraints & 
        \onlinecite{AlfCoyEhrLom-21, AlfCoyEhr-21} & (to be assessed) \\ 
        Genetic column generation (3D tests not yet avaliable)                  & \onlinecite{FriSchVoe-21}                 & 30           \\
        \hline
    \end{tabular}
\end{table*}

\section{Approximations to the SCE functional}
\label{sec_approx}

\begin{table*}
\caption{Approximate $w_\infty(\rv)$ energy densities yielding $W_\infty[\rho]$ from: $W_\infty[\rho]= \int \rho(\rv) w_\infty(\rv) \mathrm{d}\rv$.
   PC stands for point-charge plus continuum (PC) model, \cite{SeiPerKur-PRA-00} LDA stands for the local density approximation, GEA for the gradient expansion approximation,
   GGA for the generalized gradient approximation, and hPC \cite{SDGF22} stands for
   the harmonium PC based on a GGA form,\cite{PerBurErn-PRL-96} whose parameters 
   $a$ and $b$ are
   trained on the SCE energetics for the harmonium atom (for their numerical values see  Ref.~\onlinecite{SDGF22}).
   The reduced density gradient, $s$, is given by $s(\rv)=
\left| \nabla \rho (\rv)
\right| / \bigl( 2 (3\pi^2)^{1/3} \rho(\rv)^{4/3} \bigr)$.
The nonlocal radius functional (NLR) \cite{WagGor-PRA-14} approximates the strong coupling limit of the XC hole, whose depth is given by Eq.~\ref{eq:u1} and is calculated from the integrals over the spherically averaged density (Eq.~\eqref{eq:sadens}).
The shell model \cite{BahZhoErn-JCP-16} adds a positive shell to the NLR hole, and the radii of the negative and positive shell, $u_s(\rv)$ and $u_c(\rv)$, respectively, are obtained at each $\rv$ from the uniform electron gas constraint and the normalization constraint on the underlying XC hole.
The approximate $w_\infty(\rv)$ from PC-LDA, NLR and shell model are in the gauge of Eq.~\eqref{eq:wxc_def} and thereby directly approximate $w_\infty(\rv)$ of Eq.~\eqref{eq:winf}. 
}\label{tab_ap}
    \begin{tabular} {|c|c|c|}
    \hline 
    Approximation & $w_\infty(\rv)$ form & Refs. \\ \hline    
    PC-LDA & $\displaystyle -\frac{9}{10}\left(\frac{4\pi}{3}\right)^{1/3}\rho(\rv)^{1/3}$ & \onlinecite{SeiPerKur-PRA-00}    \\ \hline
    PC-GEA & $\displaystyle w_{\infty}^{\rm PC-LDA}(\rv)+9\frac{ 2^{1/3} \pi}{175}  \rho(\rv)^{1/3} s(\rv)^2$  & \onlinecite{SeiPerKur-PRA-00}   \\ \hline
    GGA (hPC) & $\displaystyle w_{\infty}^{\rm PC-LDA}(\rv)\, \frac{1+ a\, s(\rv)^2}{1+b\, s(\rv)^2}$ & \onlinecite{SDGF22} \\ \hline
    NLR & $\displaystyle -2\pi\int_0^{u_1(\rv)} \tilde{\rho}(\rv,u)u~ \dd u$  & \onlinecite{WagGor-PRA-14} \\ \hline
   shell model & $\displaystyle -2\pi\int_0^{u_s(\rv)} \tilde{\rho}(\rv,u)u~ \dd u+2\pi\int_{u_s(\rv)}^{u_c(\rv)} \tilde{\rho}(\rv,u)u~ \dd u$ & \onlinecite{BahZhoErn-JCP-16} \\ \hline
   \end{tabular}
   
\end{table*}

\begin{figure}
\includegraphics[width=\linewidth]{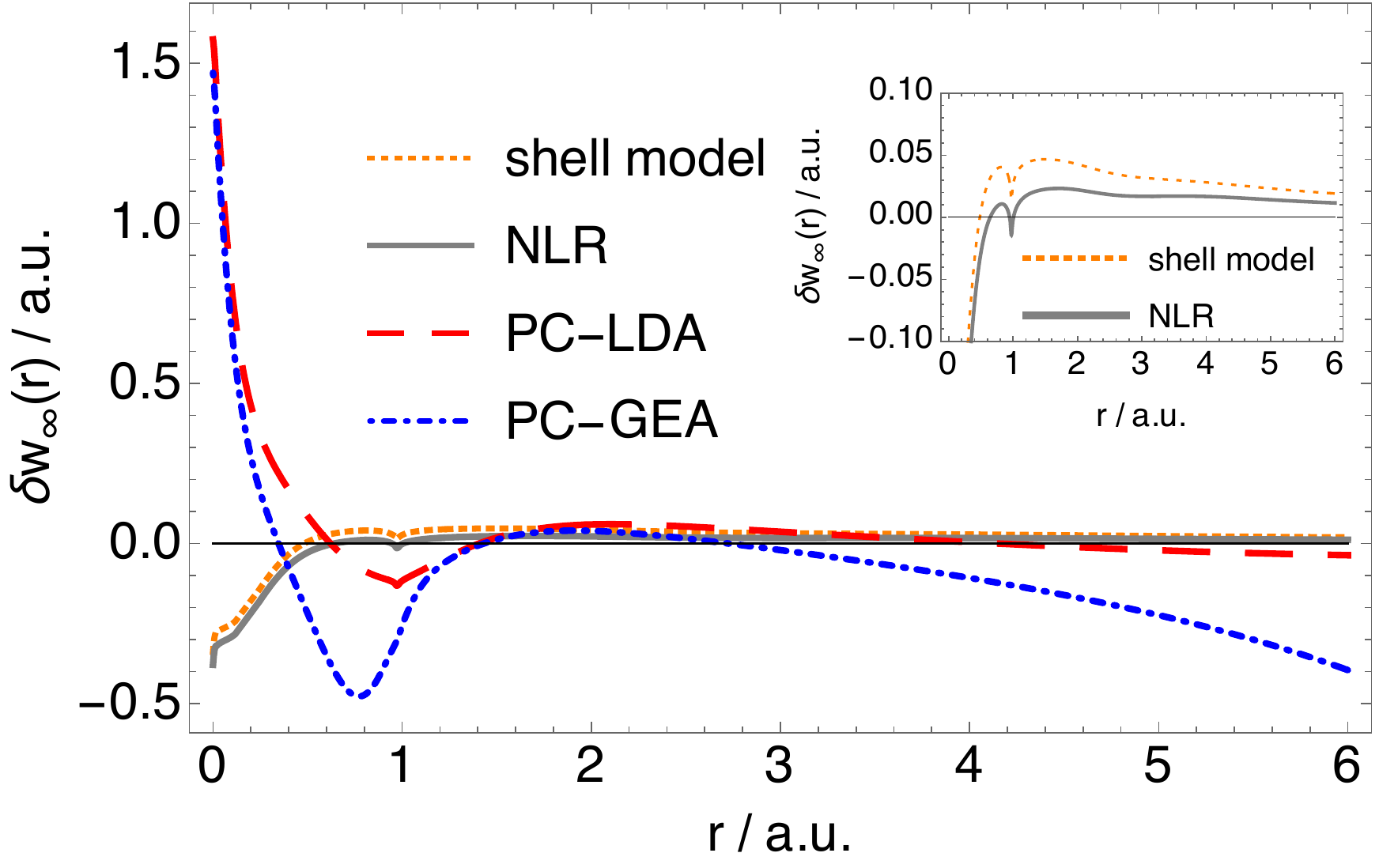}\\[2ex]
\includegraphics[width=\linewidth]{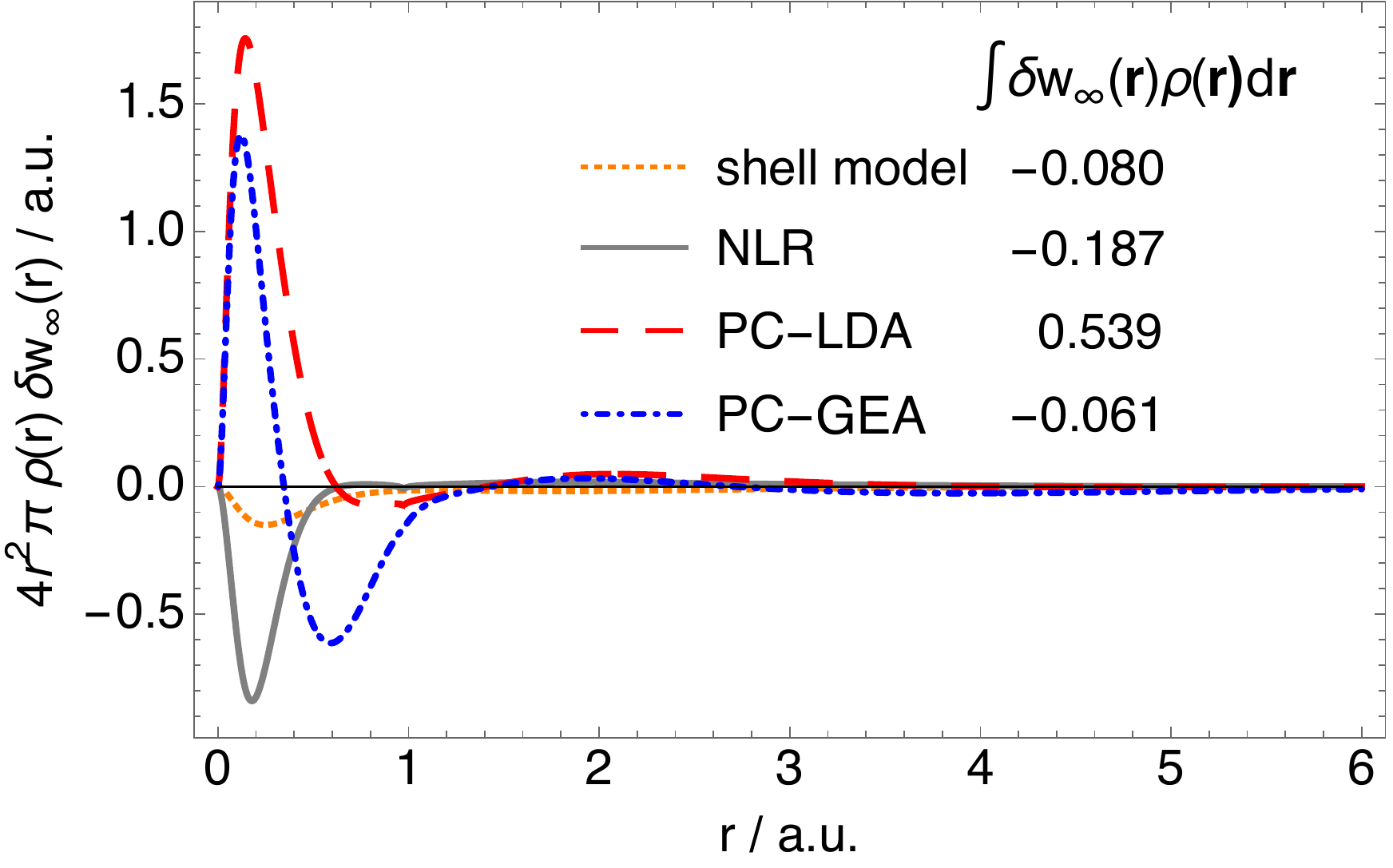}
\caption{Upper panel:  Difference between the exact (SCE) and approximate strong coupling limit energy density for the beryllium atom, $\delta w_\infty(\rv)=w_\infty(\rv)-w_\infty^{\rm model}(\rv)$, as a function of the distance from the nucleus, r / a.u. The inset in the upper panel is focusing on the error of the shell and NLR model for larger $r$. Lower panel: The quantity from the upper panel multiplied by the density and the spherical volume element.}
\label{fig_be}
\end{figure}

Existing approximations to the SCE functional are summarized in Table~\ref{tab_ap}.  
They include the 
point-charge plus continuum (PC) model of Seidl and coworkers,\cite{SeiPerKur-PRA-00} which is a gradient expansion (GEA), and the recent harmonium PC (hPC) model, which is a generalized gradient approximation (GGA).\cite{SDGF22}
The nonlocal radius functional (NLR) and the shell model retain some of the SCE non-locality \cite{WagGor-PRA-14,BahZhoErn-JCP-16} through the integrals of the spherically averaged density, which is defined as:
\begin{equation}
\tilde{\rho}(\rv,u) = \frac{1}{4\pi} \int \rho(\br + \bu) \mathrm{d}\mathbf{\Omega_{\bu} }.
\label{eq:sadens}
\end{equation}
NLR approximates the XC hole in the strong coupling limit whose depth (``nonlocal radius''), $u_1(\rv)$, is implicitly defined through the following integral, inspired by the exact SCE functional for 1D systems,
\beq
\label{eq:u1}
4 \pi \int_0^{u_1(\rv)}  u^2 \tilde{\rho}(\rv,u) \mathrm{d}u =1.
\eeq
Once $u_1(\rv)$ is computed, the energy density from the electrostatic potential of the NLR XC hole is computed, which in turn, defines $W_\infty[\rho]$ within NLR. 
The shell model is built upon NLR and makes it exact
for the SCE limit of the uniform electron gas.\cite{BahZhoErn-JCP-16} 

In Figure~\ref{fig_be}, we explore the accuracy of different SCE approximations for energy densities (see Eq.~\ref{eq:winf} below). From this figure, we can see that the shell model 
is the most accurate approximation locally. The PC-GEA model is the best performer globally here, and generally it gives a  rather accurate $W_\infty[\rho]$. 
However, the functional derivative of the PC GEA diverges in the exponentially-decaying density tails,
 \cite{FabSmiGiaDaaDelGraGor-JCTC-19,SmiCon-JCTC-20,SDGF22} making self-consistent KS calculations impossible. 
This problem is solved by turning to GGA's. \cite{SmiCon-JCTC-20,SDGF22} In particular,
the very recently proposed hPC functional\cite{SDGF22} preserves the accuracy of $W_\infty[\rho]$ from PC-GEA, while making self-consistent KS calculations possible.

\section{From SCE to practical methods}

The {\it bare} SCE functional is not directly applicable in chemistry as it over-correlates electrons.
If we take the dissociation curve of H$_2$ as an example \cite{CheFriMen-JCTC-14,VucWagMirGor-JCTC-15},
we can see that the SCE, unlike nearly all available XC approximations, dissociates the H$_2$ correctly without artificially breaking any symmetries, but predicts far too low energies around equilibrium and too short bond lengths.
For this reason,  
SCE is not directly applicable in chemistry. 
Instead one should devise smarter strategies for incorporating the SCE in an approximate XC functional.
The challenge is then to use the SCE information to equip new functionals with the ability to capture strong electronic correlations,  while maintaining the accuracy of the standard DFT for weakly and moderately correlated systems. These strategies and challenges that come along the way are discussed in the next sections. 

\subsection{Functionals via {\it global} interpolations between weak and strong coupling limit of DFT}
\label{sec_global}
XC approximations of different classes have been constructed from 
models to the global AC integrand (Eqs.~\ref{eq:ac_xc}).\cite{Ern-CPL-96,BurErnPer-CPL-97,Bec-JCP-93,MorCohYan-JCPa-06,song2021density} 
A possible way to avoid bias towards the weakly correlated regime present in nearly all XC approximations is to also include
the information from the strongly interacting limit of $W_\lambda[\rho]$.
Such an approach, called the interaction strength interpolation (ISI), where 
$W_\lambda[\rho]$ is obtained from an interpolation between its  weakly- and strongly- interacting limits, has been proposed by Seidl and coworkers.\cite{SeiPerLev-PRA-99} 
Since the ISI approach has been proposed, different interpolation forms with different input ingredients have been tested.\cite{SeiPerKur-PRL-00,SeiPerKur-PRA-00,GorVigSei-JCTC-09,LiuBur-PRA-09,ZhoBahErn-JCP-15,BahZhoErn-JCP-16,VucIroSavTeaGor-JCTC-16,VucIroWagTeaGor-PCCP-17} 
These approaches typically use the exact information from the weakly interacting limit (exact exchange and the correlation energy from the second-order Görling–Levy perturbation theory (GL2)\cite{GorLev-PRB-93}). 
Except for some proof-of-principle calculations,\cite{MalMirGieWagGor-PCCP-14,VucIroSavTeaGor-JCTC-16,KooGor-TCA-18}
the ISI scheme uses the approximate ingredients from 
the large-$\lambda$ limit ($W_\infty[\rho]$ and the next leading term described by $F^{\rm ZPE} [\rho]$) and these are
typically modeled at a semilocal level. In some cases,\cite{ZhoBahErn-JCP-15,BahZhoErn-JCP-16} the ISI forms have been tested in tandem with the $W_\infty[\rho]$ approximations that retain some of the SCE nonlocality (see Sec.~\ref{sec_approx}).

\begin{figure}
\includegraphics[width=0.45\textwidth]{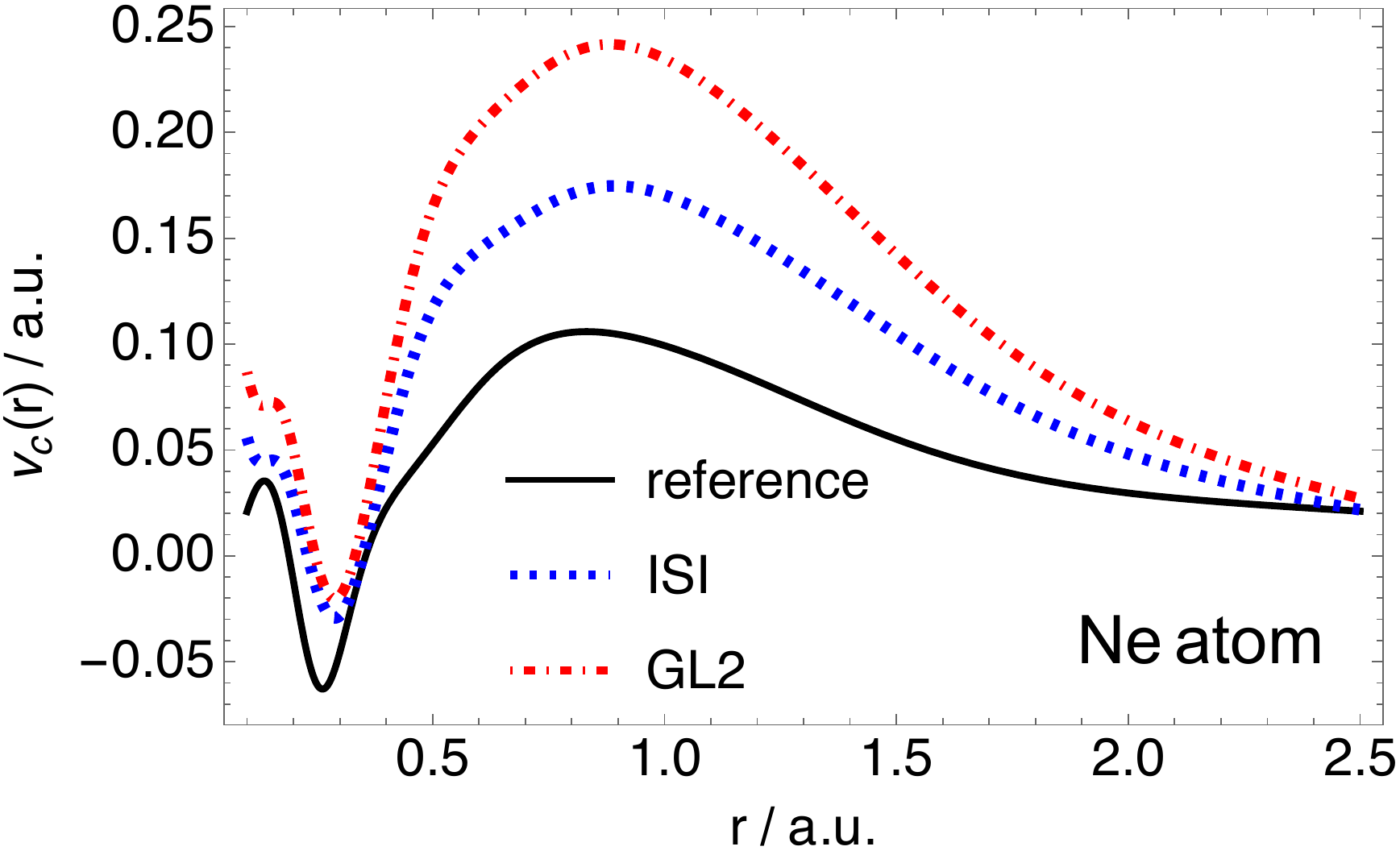}
\caption{Correlation potentials $v_c$ as a function of the distance from the nucleus ($r$) for the neon atom. The accurate correlation potential has been obtained from Quantum Monte Carlo (QMC). All the data have been taken from Ref.~\onlinecite{FabSmiGiaDaaDelGraGor-JCTC-19}. }
\label{fig:ne}
\end{figure}

A potential problem of the ISI functionals is the lack of size-consistency, which, however, can be easily corrected for interaction energies when there are no degeneracies. \cite{VucGorDelFab-JPCL-18} 
The ISI functionals have been tested on several chemical data sets and systems and they perform reasonably well for 
interaction energies (energy differences). \cite{FabGorSeiDel-JCTC-16,GiaGorDelFab-JCP-18,VucGorDelFab-JPCL-18} 
When applied in the post-SCF fashion, the ISI approach seems more promising when used in tandem with Hartree-Fock (HF) than with semilocal Kohn-Sham orbitals. 
This finding has initiated the study of the strongly interacting limit in the Hartree-Fock theory\cite{SeiGiaVucFabGor-JCP-18,DaaGroVucMusKooSeiGieGor-JCP-20} and the successes of approaches based on it will be briefly described in Sec.~\ref{sec_hf}.
Recently, the correlation potential from the ISI approach, which is needed for self-consistent ISI calculations to obtain the density and KS orbitals, has been computed. \cite{FabSmiGiaDaaDelGraGor-JCTC-19} It has been shown that it is rather accurate for a set of small
atoms 
and diatomic molecules (see Fig.~\ref{fig:ne}, where we show that the ISI correlation potential provides a substantial improvement over that from GL2 for the neon atom). \cite{FabSmiGiaDaaDelGraGor-JCTC-19}
The computed ISI correlation potentials have enabled fully self-consistent ISI calculations that have been recently reported in Ref.~\cite{SDGF22}.

\subsection{Functionals via {\it local} interpolations between weak and strong coupling limit of DFT}
\label{sec_local}

In addition to building models for the XC energy via global interpolations between the strongly- and weakly- interacting limits of $W_\lambda[\rho]$, one can also perform 
the interpolation locally (i.e. in each point of space). \cite{VucIroSavTeaGor-JCTC-16,VucIroWagTeaGor-PCCP-17}
This can be done by interpolating between the weakly and strongly interacting limits of $w_\lambda([\rho];\rv)$, which is the $\lambda$-dependent XC energy density of Eq.~\eqref{eq:wxc_def}. 
The main advantage of local interpolations\cite{VucIroSavTeaGor-JCTC-16,VucIroWagTeaGor-PCCP-17,KooGor-TCA-18} over their global counterparts is that the former are size-consistent by construction if the interpolation ingredients are size-consistent.\cite{GorSav-JPCS-08,Sav-CP-09}  Thereby, local interpolations, unlike their global counterparts, do not require the size-consistency correction .\cite{VucLevGor-JCP-17} 

As mentioned earlier, there is no unique definition for $w_\lambda([\rho];\rv)$. 
Ref.~\onlinecite{VucLevGor-JCP-17} explored the suitability of different definitions of the $\lambda$-dependent energy densities and it has been found that the energy densities definition of Eq.~\eqref{eq:wxc_def} ('electrostatic potential of the XC hole') is the best choice so far in this context. 
Within this definition, $w_\lambda([\rho];\rv)$ reduces to the exact exchange energy density when $\lambda = 0$, whereas in the $\lambda \to \infty$ (within the SCE formulation), it is defined in terms of the co-motion functions \cite{MirSeiGor-JCTC-12}: 
\begin{align}
	w_\infty([\rho];\rv)= \sum_{i=2}^{N}\frac{1}{\left | \rv-\fv_i(\rv) \right|}  -\frac{1}{2}v_H(\rv),
\label{eq:winf}
	\end{align}
where $v_H(\rv)$ is the Hartree potential. In addition to these two, a closed form expression for the local initial slope for $w_\lambda([\rho];\rv)$ has been derived in Ref.~\onlinecite{VucIroSavTeaGor-JCTC-16} from second order perturbation theory. 

The accuracy of different local interpolation forms have been tested with both exact \cite{VucIroSavTeaGor-JCTC-16,KooGor-TCA-18} and approximate~\cite{VucIroSavTeaGor-JCTC-16,ZhoBahErn-JCP-15,BahZhoErn-JCP-16} ingredients. Relative to the global interpolations, local interpolations typically give improved results for tested small chemical systems, \cite{VucIroSavTeaGor-JCTC-16} but usually do not fix the failures of global interpolations.\cite{VucIroWagTeaGor-PCCP-17} Nevertheless, the accuracy of XC functionals based on the local interpolation is still underexplored. This local interpolation framework can also be used to improve the latest XC approximations, such as the deep learned {\it local hybrids},\cite{fn_dm21} especially when it comes to the treatment of strong electronic correlations. 

\subsection{Fully nonlocal multiple radii functional - inspired by the exact SCE form}

The mathematical form of the SCE functional has inspired new fully nonlocal approximations, called the multiple radii functional (MRF). \cite{VucGor-JPCL-17,VJCTC19,VG19} MRF approximates the XC energy denities of Eq.~\eqref{eq:wxc_def} at arbitrary $\lambda$ in the following way:
\begin{equation}
w_\lambda^{\rm MRF}(\rv)= \frac{1}{2} \sum_{i=2}^{N} \frac{1}{R_i^{\lambda}([\rho];\rv)}-\frac{1}{2}v_H(\rv). 
\label{eq:mrf1}
\end{equation}
Equation~\ref{eq:mrf1} can be thought as the generalization of Eq.~\eqref{eq:winf}, where starting from a reference electron at $\rv$, the remaining electrons are assigned effective radii or distances from $\rv$. The radii are then constructed from the integrals over the spherically averaged density and are implicitly defined by
\begin{equation}
4 \pi \int_0^{R_i^{\lambda}(\rv) }  u^2 \tilde{\rho}(\rv,u) \mathrm{d}u = i - 1 + \sigma_i^{\lambda}(\rv),
\end{equation}
where $\sigma_i^{\lambda}(\rv)$ is the so-called {\it fluctuation function}. 
The construction of the XC functional within MRF essentially reduces to building $\sigma_i^{\lambda}(\rv)$ , 
and  already very simple forms for $\sigma_i^{\lambda=1}(\rv)$ yield very accurate 
atomic $w_\lambda^{\rm MRF}(\rv)$ at the physical regime for atoms, while also accurately capturing the physics of stretched bonds. This shows that the forms inspired by the SCE can work for the physical regime if properly re-scaled. 
Furthermore, despite its full non-locality, the cost of MRF is  $O(N^3)$ within seminumerical schemes.\cite{BahmannKaupp2015} 

By construction, MRF has very appealing properties: (1) it gives XC energies in the gauge of Eq.~\eqref{eq:wxc_def} making it highly suitable to be used in the local interpolations described in Sec.~\ref{sec_local}; (2) these energy densities have the correct asymptotic behavior; (3) MRF captures the physics of bond breaking; (4) it is fully nonlocal so it can better describe the physics of strong electronic correlations that the usual semilocal DFT functionals; (5) its form is universal and does not change as dimensionality/interactions between particle changes as demonstrated in Ref.~\onlinecite{VG19}.
All these features of MRF and its flexibility make it very promising for building the next-generation of DFT approximations. There are ongoing efforts to 
transform these appealing features into robust XC functionals by developing improved MRF forms and efficiently implementing the MRF package into standard quantum-chemical codes.

\subsection{Other applications of SCE: Lower bounds to XC energies and correlation indicators}

Besides being used to build XC approximations, the SCE approach has also  proven very useful in understanding general features of the exact XC functional and the nature of electronic correlations.
For example, the SCE limit is directly connected to the Lieb-Oxford (LO) inequality, \cite{Lie-PLA-79,LieOxf-IJQC-81} a key exact property used in the construction of XC approximations.\cite{Per-INC-91,SunRuzPer-PRL-15} 
The LO inequality limits the value of the XC energy by 
bounding from below the AC integrand of Eq.~\eqref{eq:w1}:
\beq
W_\lambda [\rho] \geq - C_{\rm LO} \int  \rho ^{4/3} (\rv) \mathrm{d} \rv,
\label{eq:lo}
\eeq
where the optimal $C_{\rm LO}$ is rigorously known to be between 1.4442 and 1.5765. \cite{LewLieSei-PRB-19,CotPet-arxiv-17,LLS22}
Since $W_\lambda [\rho]$ monotonically decreases with $\lambda$, $W_\infty [\rho]$ will be the smallest value for the l.h.s.~of Eq.~\eqref{eq:lo}. 
Thus, finding lower bounds for the optimal constant $C_{\rm LO}$ is equivalent to searching for densities $\rho$ that maximize the ratio between $W_\infty [\rho]$ and $-\int \rho ^{4/3} (\rv) \mathrm{d} \rv$, \cite{RasSeiGor-PRB-11,SeiVucGor-MP-16} a procedure that has been applied to both the optimal $C_{\rm LO}$ for the general case and to the one for a specific number of electrons $N$. \cite{RasSeiGor-PRB-11,SeiVucGor-MP-16,VucLevGor-JCP-17,SBKG22} 
An approach to tighten the lower bound to correlation energies for a given density has been also proposed  
by combing the adiabatic connection interpolation described in Sec.~\ref{sec_global} and the SCE energies.\cite{VucIroWagTeaGor-PCCP-17}

In addition to provide tightened lower bounds for the XC energies, the SCE has also been used to define correlation indicators that quantify the ratio between {\it dynamical} and {\it static} correlation in a given system.\cite{VucIroWagTeaGor-PCCP-17}
This idea has been also generalized to local indicators, enabling to visualise
the interplay of dynamical and static correlation
at different points in space.\cite{VucIroWagTeaGor-PCCP-17} 

\begin{figure}
\includegraphics[width=0.5\textwidth]{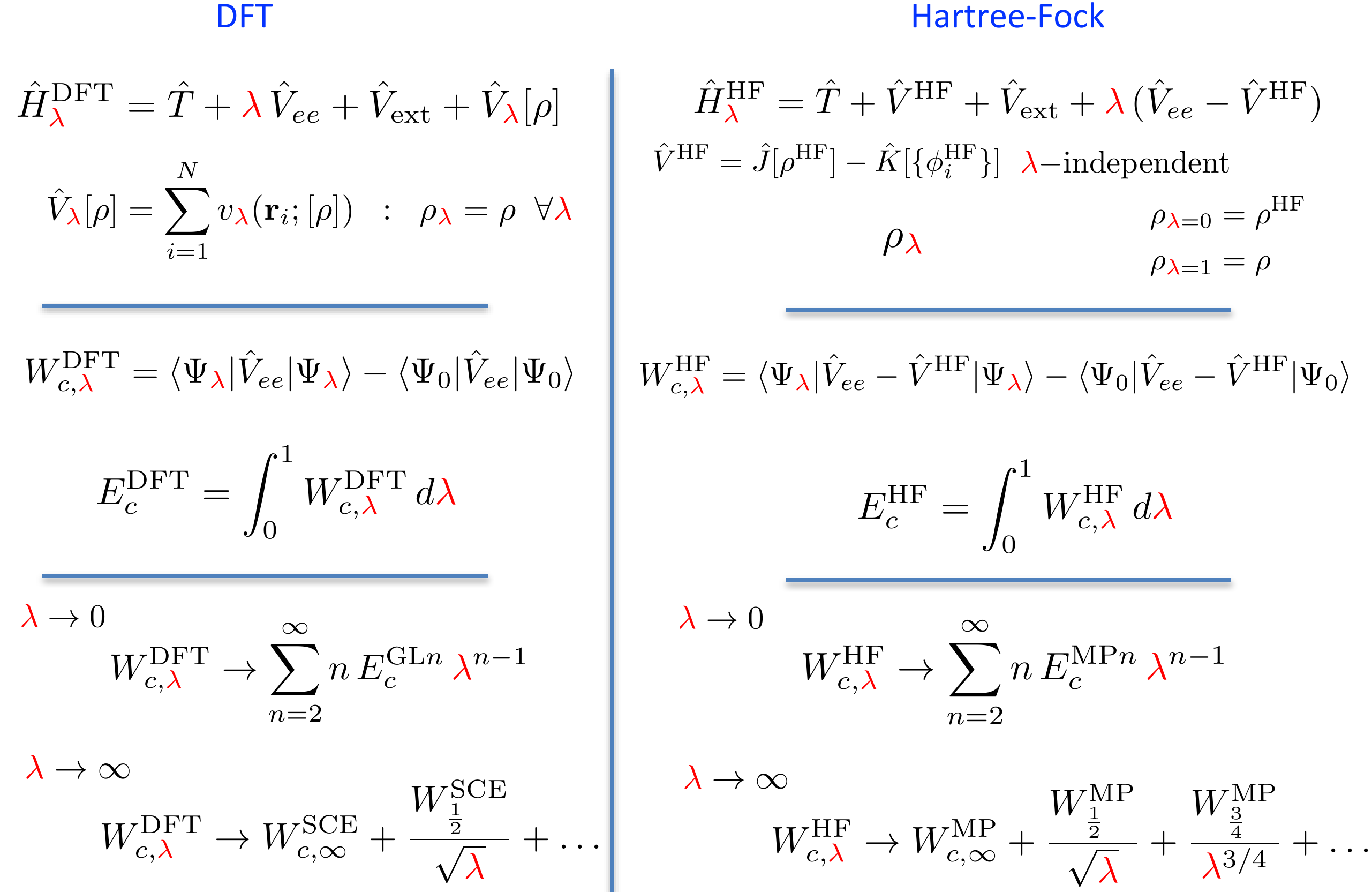}
\caption{Two different adiabatic connections, linking the physical ($\lambda=1$) system to either the KS or the Hartree-Fock determinant ($\lambda=0$). Left: the density-fixed adiabatic connection of KS DFT (see Sec.~\ref{sec_theory}). The Hamiltonian $\hat{H}_\lambda^{\rm DFT}$ corresponds to Eq.~\eqref{eq.LLalpha}, with the one-body potential $v_\lambda(\rv)$ enforcing the constraint $\Psi_\lambda\mapsto \rho$, where $\rho$ is the density of the physical system. 
The correlation part of the adiabatic connection integrand $W_{c,\lambda}^{\rm DFT}$ is equal to $W_\lambda[\rho]$ of Eq.~\eqref{eq:w1} minus $W_{\lambda=0}[\rho] = E_x[\rho]$. 
Small and large-$\lambda$
expansions for $W_{c,\lambda}^{\rm DFT}$ are also shown.
Right: the adiabatic connection that has the M{\o}ller-Plesset (MP) series as small perturbation expansion, considered in Sec.~\ref{sec_hf}. The Hamiltonian $\hat{H}_\lambda^{\rm HF}$ contains  $\hat{J}[\rho^{\rm HF}]$ and $\hat{K}[\{\phi_i^{\rm HF}\}]$, which are the standard Hartree and exchange HF operators, respectively.
The $\lambda$-dependent $\Psi$ minimizing $\hat{H}_\lambda^{\rm HF}$ has a density that changes with $\lambda$: at $\lambda=0$ is equal to the HF density, while at $\lambda=1$ is equal to the physical density. The expectation $W_{c,\lambda}^{\rm HF}$ is the AC integrand defining the correlation energy in HF theory.
Small and large-$\lambda$
expansions for $W_{c,\lambda}^{\rm HF}$ are also shown.
The operator $\hat{V}_{\rm ext}$ is the external (nuclear) potential.
}
\label{fig:adiabs}
\end{figure}

\subsection{Going beyond DFT -- large-$\lambda$ limits in the M{\o}ller-Plesset adiabatic connection}
\label{sec_hf}

\begin{figure}
\includegraphics[width=\columnwidth]{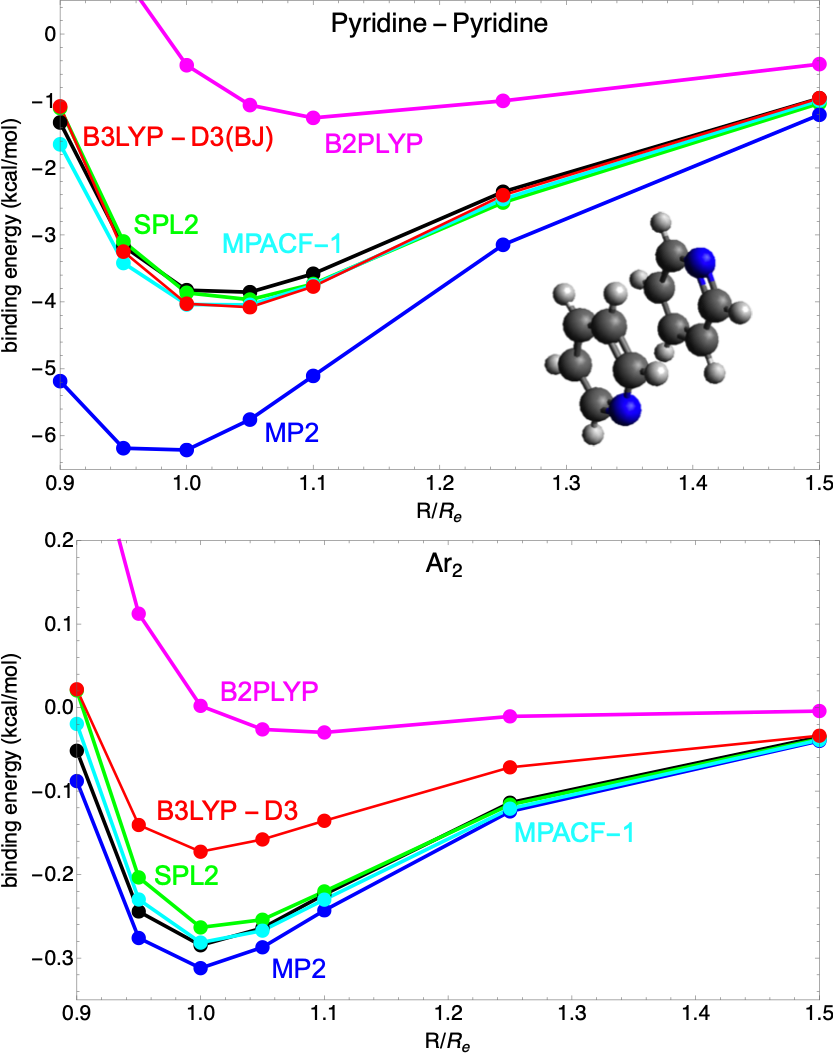}
\caption{Dissociation curves of the Pyridine (top) and Argon dimers (bottom) obtained from MP2, SPL2, MPACF-1, B3LYP-D3, and B2PLYP with CCSD(T) (black line) as a reference.}
\label{fig:Arg}
\end{figure}

The DFT AC introduced in Sec.~\ref{sec_xc}, whose large $\lambda$ limit is the focus of this paper, defines the correlation energy in KS DFT. In a more traditional quantum-chemical sense, the correlation energy is defined as the difference between the true and Hartree-Fock (HF) energy. 
An exact expression for this correlation energy is given by
the M{\o}ller-Plesset adiabatic connection (MPAC),\cite{SeiGiaVucFabGor-JCP-18} which connects the HF and physical state and has the Møller--Plesset perturbation series as weak-interaction
expansion. This AC is summarized and compared with the one of KS DFT used in the rest of this work in Fig.~\ref{fig:adiabs}. The large-$\lambda$ limit of the MP AC has been recently shown\cite{SeiGiaVucFabGor-JCP-18,DaaGroVucMusKooSeiGieGor-JCP-20} and to be determined by functionals of the HF density with a clear physical meaning.
Inequalities between the large-$\lambda$ leading terms of the two AC's have been also established.\cite{SeiGiaVucFabGor-JCP-18,DaaGroVucMusKooSeiGieGor-JCP-20}

The MP AC theory has been used to construct 
  a predictor for the accuracy of MP2  for noncovalent interactions.\cite{VucFabGorBur-JCTC-20}
Methods that are based on the interpolation between the small and large $\lambda$ limits of the MP AC have been also developed.  \cite{DaaFabDelGorVuc-JPCL-21} They are analogous to the ISI methods outlined in Sec.~\ref{sec_global}, which are used in the DFT context. It has been shown that the these interpolation methods for the MP AC give very accurate results for noncovalent interactions. \cite{DaaFabDelGorVuc-JPCL-21} We illustrate this in Fig.~\ref{fig:Arg}, where we compare reference [CCSD(T)] to approximate dissociation curves from these interpolation approximations for the pyridine and argone dimers.
The curves labelled SPL2 and MPACF-1 correspond to two new global interpolations forms\cite{DaaFabDelGorVuc-JPCL-21} constructed by adding more flexibility and empirical parameters to the existing interpolation forms used in DFT\cite{SeiPerLev-PRA-99,GorVigSei-JCTC-09} to capture the known exact features of the MP AC. For both of these noncovalently bound dimers, and for many other cases,\cite{DaaFabDelGorVuc-JPCL-21} SPL2 and MPACF-1 show an excellent performance without using dispersion corrections. In general, they substantially improve over MP2 for noncovalent interactions, and are either on par with -- or also improve -- dispersion corrected (double) hybrids \cite{DaaFabDelGorVuc-JPCL-21}.

\section{Conclusions and Outlook}

Here we have reviewed the most important topics that the strongly interacting limit of DFT brings into focus.
We have analyzed the development of different aspects of the underling rigorous theory connecting DFT and optimal transport, and discussed 
how the SIL formulation 
influenced the development of different
methods in 
DFT and beyond. Although this limit does not describe the physical regime, its mathematical structure contains essential elements pointing towards the real physics happening in molecular systems with strong correlations, whose description is  one of the key unsolved problems in DFT.
Thus, in the years and decades to come 
it will be very interesting to see how much the SIL ideas, formulations and ensuing practical methods 
will be used to solve the strong correlation problem and to build the next generation of DFT methods. In particular, the new ingredients appearing in this limit can be used as new features to machine learn the XC functional.\cite{Kalita2021,fn_dm21,Nagai2020} 

\section*{Conflicts of Interest}   
All the authors declare to have no conflict of interest.

\section*{Data Sharing}   
Data sharing is not applicable to this article as no new data were created or analyzed in this study.

\section*{Acknowledgement}    
SV acknowledges funding from the Marie Sklodowska-Curie grant 101033630 (EU’s Horizon 2020 programme). AG acknowledges support of his research by the Canada Research Chairs Program and Natural Sciences and Engineering Research Council of Canada (NSERC), funding reference number RGPIN-2022-05207.
HB acknowledges funding by the Deutsche Forschungsgemeinschaft (DFG, German Research Foundation) – project no. 418140043. TJD and PG-G were supported by the Netherlands Organisation for Scientific Research (NWO) under Vici grant 724.017.001. GF was partially supported by the Deutsche Forschungsgemeinschaft (DFG, German Research Foundation) through CRC 109.

\bibliographystyle{vancouver}
\bibliography{biblio,bib_clean}

\begin{thebibliography}{10}

\bibitem{Bur-JCP-12}
Burke K.
\newblock Perspective on density functional theory.
\newblock J Chem Phys. 2012;136(15):150901.

\bibitem{PGB15}
Pribram-Jones A, Gross DA, Burke K.
\newblock DFT: A Theory Full of Holes?
\newblock Annual Review of Physical Chemistry. 2015;66(1):283--304.
\newblock Available from:
  \url{http://www.annualreviews.org/doi/abs/10.1146/annurev-physchem-040214-121420}.

\bibitem{MH17}
Mardirossian N, Head-Gordon M.
\newblock Thirty years of density functional theory in computational chemistry:
  an overview and extensive assessment of 200 density functionals.
\newblock Molecular Physics. 2017;115(19):2315--2372.

\bibitem{GHBE17}
Goerigk L, Hansen A, Bauer C, Ehrlich S, Najibi A, Grimme S.
\newblock A look at the density functional theory zoo with the advanced GMTKN55
  database for general main group thermochemistry, kinetics and noncovalent
  interactions.
\newblock Physical Chemistry Chemical Physics. 2017;19(48):32184--32215.

\bibitem{VT20}
Verma P, Truhlar DG.
\newblock Status and challenges of density functional theory.
\newblock Trends in Chemistry. 2020;2(4):302--318.

\bibitem{Sim2022}
Sim E, Song S, Vuckovic S, Burke K.
\newblock Improving Results by Improving Densities: Density-Corrected Density
  Functional Theory.
\newblock Journal of the American Chemical Society. 2022
  Apr;144(15):6625--6639.
\newblock Available from: \url{https://doi.org/10.1021/jacs.1c11506}.

\bibitem{Sherrill2020}
Sherrill CD, Manolopoulos DE, Mart{\'{\i}}nez TJ, Michaelides A.
\newblock Electronic structure software.
\newblock The Journal of Chemical Physics. 2020 Aug;153(7):070401.
\newblock Available from: \url{https://doi.org/10.1063/5.0023185}.

\bibitem{SunRuzPer-PRL-15}
Sun J, Ruzsinszky A, Perdew JP.
\newblock Strongly constrained and appropriately normed semilocal density
  functional.
\newblock Phys Rev Lett. 2015;115(3):036402.

\bibitem{Kalita2021}
Kalita B, Li L, McCarty RJ, Burke K.
\newblock Learning to Approximate Density Functionals.
\newblock Accounts of Chemical Research. 2021 Feb;54(4):818--826.
\newblock Available from: \url{https://doi.org/10.1021/acs.accounts.0c00742}.

\bibitem{fn_dm21}
Kirkpatrick J, McMorrow B, Turban DH, Gaunt AL, Spencer JS, Matthews AG, et~al.
\newblock Pushing the frontiers of density functionals by solving the
  fractional electron problem.
\newblock Science. 2021;374(6573):1385--1389.

\bibitem{Nagai2020}
Nagai R, Akashi R, Sugino O.
\newblock Completing density functional theory by machine learning hidden
  messages from molecules.
\newblock npj Computational Materials. 2020 May;6(1).
\newblock Available from: \url{https://doi.org/10.1038/s41524-020-0310-0}.

\bibitem{PerSch-AIP-01}
Perdew JP, Schmidt K.
\newblock Jacob’s ladder of density functional approximations for the
  exchange-correlation energy.
\newblock In: AIP Conference Proceedings. vol. 577. AIP; 2001. p. 1--20.

\bibitem{Ham-SCI-17}
Hammes-Schiffer S.
\newblock A conundrum for density functional theory.
\newblock Science. 2017;355(6320):28--29.

\bibitem{CohMorYan-CR-12}
Cohen AJ, Mori-S\'anchez P, Yang W.
\newblock Chem Rev. 2012;112:289.

\bibitem{Sei-PRA-99}
Seidl M.
\newblock Phys Rev A. 1999;{60}:4387.

\bibitem{SeiGorSav-PRA-07}
Seidl M, Gori-Giorgi P, Savin A.
\newblock Phys Rev A. 2007;{75}:042511.

\bibitem{GorVigSei-JCTC-09}
Gori-Giorgi P, Vignale G, Seidl M.
\newblock J Chem Theory Comput. 2009;{5}:743.

\bibitem{ButDepGor-PRA-12}
Buttazzo G, De~Pascale L, Gori-Giorgi P.
\newblock Optimal-transport formulation of electronic density-functional
  theory.
\newblock Phys Rev A. 2012 6;85(6):062502.

\bibitem{FriGerGor-arxiv-22}
Friesecke G, Gerolin A, Gori-Giorgi P.
\newblock The strong-interaction limit of density functional theory.
\newblock arXiv preprint arXiv:220209760. 2022.

\bibitem{DaaFabDelGorVuc-JPCL-21}
Daas TJ, Fabiano E, Della~Sala F, Gori-Giorgi P, Vuckovic S.
\newblock Noncovalent interactions from models for the m{\o}ller--plesset
  adiabatic connection.
\newblock The journal of physical chemistry letters. 2021;12(20):4867--4875.

\bibitem{Lev-PNAS-79}
Levy M.
\newblock Proc Natl Acad Sci USA. 1979;76:6062.

\bibitem{Lie-IJQC-83}
Lieb EH.
\newblock Int J Quantum Chem. 1983;{24}:24.

\bibitem{LanPer-SSC-75}
Langreth DC, Perdew JP.
\newblock Solid State Commun. 1975;{17}:1425.

\bibitem{GunLun-PRB-76}
Gunnarsson O, Lundqvist BI.
\newblock Exchange and correlation in atoms, molecules, and solids by the
  spin-density-functional formalism.
\newblock Phys Rev B. 1976;{13}:4274.

\bibitem{BurCruLam-JCP-98}
Burke K, Cruz FG, Lam KC.
\newblock J Chem Phys. 1998;109:8161.

\bibitem{VucIroSavTeaGor-JCTC-16}
Vuckovic S, Irons TJP, Savin A, Teale AM, Gori-Giorgi P.
\newblock Exchange--correlation functionals via local interpolation along the
  adiabatic connection.
\newblock J Chem Theory Comput. 2016;12(6):2598--2610.

\bibitem{VucGor-JPCL-17}
Vuckovic S, Gori-Giorgi P.
\newblock Simple Fully Nonlocal Density Functionals for Electronic Repulsion
  Energy.
\newblock The Journal of Physical Chemistry Letters. 2017 6;8(13):2799--2805.
\newblock PMID: 28581751.

\bibitem{VucLevGor-JCP-17}
Vuckovic S, Levy M, Gori-Giorgi P.
\newblock Augmented potential, energy densities, and virial relations in the
  weak-and strong-interaction limits of DFT.
\newblock The Journal of Chemical Physics. 2017;147(21):214107.

\bibitem{CotFriKlu-CPAM-13}
Cotar C, Friesecke G, Kl\"uppelberg C.
\newblock Density Functional Theory and Optimal Transportation with Coulomb
  Cost.
\newblock Comm Pure Appl Math. 2013;66:548--99.

\bibitem{CotFriKlu-ARMA-18}
Cotar C, Friesecke G, Kl{\"u}ppelberg C.
\newblock Smoothing of transport plans with fixed marginals and rigorous
  semiclassical limit of the Hohenberg--Kohn functional.
\newblock Arch Ration Mech An. 2018 6;228(3):891--922.

\bibitem{Lew-CRM-18}
Lewin M.
\newblock Semi-classical limit of the {L}evy--{L}ieb functional in {D}ensity
  {F}unctional {T}heory.
\newblock C R Math. 2018 3;356(4):449--455.

\bibitem{LevPer-PRA-85}
Levy M, Perdew JP.
\newblock Phys Rev A. 1985;32:2010.

\bibitem{LevPer-PRB-93}
Levy M, Perdew JP.
\newblock Phys Rev B. 1993;{48}:11638.

\bibitem{ColDepDim-CJM-15}
Colombo M, De~Pascale L, Di~Marino S.
\newblock Multimarginal Optimal Transport Maps for One-dimensional Repulsive
  Costs.
\newblock Canad J Math. 2015 5;67:350--368.

\bibitem{MirSeiGor-JCTC-12}
Mirtschink A, Seidl M, Gori-Giorgi P.
\newblock Energy densities in the strong-interaction limit of density
  functional theory.
\newblock J Chem Theory Comput. 2012;8(9):3097--3107.

\bibitem{MalGor-PRL-12}
Malet F, Gori-Giorgi P.
\newblock Phys Rev Lett. 2012;109:246402.

\bibitem{MalMirCreReiGor-PRB-13}
Malet F, Mirtschink A, Cremon JC, Reimann SM, Gori-Giorgi P.
\newblock Phys Rev B. 2013;87:115146.

\bibitem{ColStr-M3AS-15}
Colombo M, Stra F.
\newblock Counterexamples in multimarginal optimal transport with Coulomb cost
  and spherically symmetric data.
\newblock Mathematical Models and Methods in Applied Sciences.
  2016;26(06):1025--1049.

\bibitem{SeiDiMGerNenGieGor-arxiv-17}
Seidl M, Di~Marino S, Gerolin A, Nenna L, Giesbertz KJ, Gori-Giorgi P.
\newblock The strictly-correlated electron functional for spherically symmetric
  systems revisited.
\newblock arXiv preprint arXiv:170205022. 2017.

\bibitem{MenLin-PRB-13}
Mendl CB, Lin L.
\newblock Kantorovich dual solution for strictly correlated electrons in atoms
  and molecules.
\newblock Phys Rev B. 2013;87:125106.

\bibitem{VucWagMirGor-JCTC-15}
Vuckovic S, Wagner LO, Mirtschink A, Gori-Giorgi P.
\newblock Hydrogen Molecule Dissociation Curve with Functionals Based on the
  Strictly Correlated Regime.
\newblock J Chem Theory Comput. 2015;11(7):3153--3162.

\bibitem{ColDMaStra-arxiv-21}
Colombo M, Di~Marino S, Stra F.
\newblock First order expansion in the semiclassical limit of the Levy-Lieb
  functional.
\newblock arXiv preprint arXiv:210606282. 2021.

\bibitem{GorSeiVig-PRL-09}
Gori-Giorgi P, Seidl M, Vignale G.
\newblock Phys Rev Lett. 2009;{103}:166402.

\bibitem{GroKooGieSeiCohMorGor-JCTC-17}
Grossi J, Kooi DP, Giesbertz KJH, Seidl M, Cohen AJ, Mori-S{\'a}nchez P, et~al.
\newblock Fermionic statistics in the strongly correlated limit of Density
  Functional Theory.
\newblock J Chem Theory Comput. 2017 11;13(12):6089--6100.

\bibitem{FriSchVoe-21}
Friesecke G, Schulz AS, V\"ogler D.
\newblock Genetic column generation: Fast computation of high-dimensional
  multi-marginal optimal transport problems.
\newblock to appear in SIAM J Sci Comp, arXiv preprint: arXiv:210312624. 2021.

\bibitem{linHoCutJor-arXiv-19}
Lin T, Ho N, Cuturi M, Jordan MI.
\newblock On the Complexity of Approximating Multimarginal Optimal Transport.
\newblock arXiv preprint arXiv:191000152. 2019.

\bibitem{AltBoi-arXiv-20}
Altschuler J, Boix-Adser\`a E.
\newblock Polynomial-time algorithms for {M}ultimarginal {O}ptimal {T}ransport
  problems with structure.
\newblock arXiv:200803006v1. 2020.

\bibitem{AltBoi-Dis-21}
Altschuler JM, Boix-Adsera E.
\newblock Hardness results for multimarginal optimal transport problems.
\newblock Discrete Optimization. 2021;42:100669.

\bibitem{Vuc-Thesis-17}
Vuckovic S.
\newblock Fully Nonlocal Exchange-Correlation Functionals from the
  Strongcoupling limit of Density Functional Theory.
\newblock PhD thesis. 2017.

\bibitem{SeiVucGor-MP-16}
Seidl M, Vuckovic S, Gori-Giorgi P.
\newblock Challenging the Lieb--Oxford bound in a systematic way.
\newblock Mol Phys. 2016;114:1076--1085.

\bibitem{CheFriMen-JCTC-14}
Chen H, Friesecke G, Mendl CB.
\newblock Numerical Methods for a {K}ohn-{S}ham density functional model based
  on optimal transport.
\newblock J Chem Theory Comput. 2014;10:4360--4368.

\bibitem{BenCarCutNenPey-SIAM-15}
Benamou JD, Carlier G, Cuturi M, Nenna L, Peyr{\'e} G.
\newblock Iterative bregman projections for regularized transportation
  problems.
\newblock SIAM J on Sci Comput. 2015;37(2):A1111--A1138.

\bibitem{BenCarNen-SMCISE-16}
Benamou JD, Carlier G, Nenna L.
\newblock Splitting Methods in Communication, Imaging, Science, and
  Engineering.

\bibitem{DMaGerNen-TOOAS-17}
Di~Marino S, Gerolin A, Nenna L.
\newblock Optimal Transport for Repulsive costs.
\newblock Topological Optimization and Optimal Transport In the Applied
  Sciences. 2017.

\bibitem{DMaGer-JSC-20}
Di~Marino S, Gerolin A.
\newblock An {O}ptimal {T}ransport approach for the {S}chr{\"o}dinger bridge
  problem and convergence of {S}inkhorn algorithm.
\newblock Journal of Scientific Computing. 2020;85(27).

\bibitem{GerGroGor-JCTC-20}
Gerolin A, Grossi J, Gori-Giorgi P.
\newblock Kinetic correlation functionals from the entropic regularisation of
  the strictly-correlated electrons problem.
\newblock Journal of Chemical Theory and Computation. 2019;16(1):488--498.

\bibitem{KhoYin-SIAMJSC-19}
Khoo Y, Ying L.
\newblock Convex Relaxation Approaches for Strictly Correlated Density
  Functional Theory.
\newblock SIAM J Sci Comput. 2019;41(4):B773--B795.
\newblock Available from: \url{https://doi.org/10.1137/18M1207478}.

\bibitem{AlfCoyEhrLom-21}
Alfonsi A, Coyaud R, Ehrlacher V, Lombardi D.
\newblock Approximation of optimal transport problems with marginal moments
  constraints.
\newblock Math Comp. 2021;90:689--737.

\bibitem{AlfCoyEhr-21}
Alfonsi A, Coyaud R, Ehrlacher V.
\newblock Constrained overdamped Langevin dynamics for symmetric multimarginal
  optimal transportation.
\newblock arXiv preprint: arXiv:210203091. 2021.

\bibitem{SeiPerKur-PRA-00}
Seidl M, Perdew JP, Kurth S.
\newblock Phys Rev A. 2000;{62}:012502.

\bibitem{SDGF22}
{\'S}miga S, Della~Sala F, Gori-Giorgi P, Fabiano E.
\newblock Self-consistent Kohn-Sham calculations with adiabatic connection
  models.
\newblock arXiv preprint arXiv:220211531. 2022.

\bibitem{PerBurErn-PRL-96}
Perdew JP, Burke K, Ernzerhof M.
\newblock Generalized Gradient Approximation Made Simple.
\newblock Phys Rev Lett. 1996;77:3865.

\bibitem{WagGor-PRA-14}
Wagner LO, Gori-Giorgi P.
\newblock Electron avoidance: A nonlocal radius for strong correlation.
\newblock Phys Rev A. 2014 11;90(5):052512.

\bibitem{BahZhoErn-JCP-16}
Bahmann H, Zhou Y, Ernzerhof M.
\newblock The shell model for the exchange-correlation hole in the
  strong-correlation limit.
\newblock J Chem Phys. 2016;145(12):124104.

\bibitem{FabSmiGiaDaaDelGraGor-JCTC-19}
Fabiano E, Śmiga S, Giarrusso S, Daas TJ, Della~Sala F, Grabowski I, et~al.
\newblock Investigation of the Exchange-Correlation Potentials of Functionals
  Based on the Adiabatic Connection Interpolation.
\newblock J Chem Theory Comput. 2019;15(2):1006--1015.
\newblock Available from: \url{https://doi.org/10.1021/acs.jctc.8b01037}.

\bibitem{SmiCon-JCTC-20}
Smiga S, Constantin LA.
\newblock Modified Interaction-Strength Interpolation Method as an Important
  Step toward Self-Consistent Calculations.
\newblock Journal of chemical theory and computation. 2020;16(8):4983--4992.

\bibitem{Ern-CPL-96}
Ernzerhof M.
\newblock Construction of the adiabatic connection.
\newblock Chem Phys Lett. 1996;{263}:499.

\bibitem{BurErnPer-CPL-97}
Burke K, Ernzerhof M, Perdew JP.
\newblock The adiabatic connection method: A non-empirical hybrid.
\newblock Chem Phys Lett. 1997;{265}:115.

\bibitem{Bec-JCP-93}
Becke AD.
\newblock Density-functional thermochemistry. III. The role of exact exchange.
\newblock J Chem Phys. 1993;98:5648.

\bibitem{MorCohYan-JCPa-06}
Mori-Sanchez P, Cohen AJ, Yang WT.
\newblock J Chem Phys. 2006;{124}:091102.

\bibitem{song2021density}
Song S, Vuckovic S, Sim E, Burke K.
\newblock Density sensitivity of empirical functionals.
\newblock The journal of physical chemistry letters. 2021;12(2):800--807.

\bibitem{SeiPerLev-PRA-99}
Seidl M, Perdew JP, Levy M.
\newblock Phys Rev A. 1999;{59}:51.

\bibitem{SeiPerKur-PRL-00}
Seidl M, Perdew JP, Kurth S.
\newblock Phys Rev Lett. 2000;{84}:5070.

\bibitem{LiuBur-PRA-09}
Liu ZF, Burke K.
\newblock Phys Rev A. 2009;{79}:064503.

\bibitem{ZhoBahErn-JCP-15}
Zhou Y, Bahmann H, Ernzerhof M.
\newblock Construction of exchange-correlation functionals through
  interpolation between the non-interacting and the strong-correlation limit.
\newblock J Chem Phys. 2015 9;143(12):124103.

\bibitem{VucIroWagTeaGor-PCCP-17}
Vuckovic S, Irons TJP, Wagner LO, Teale AM, Gori-Giorgi P.
\newblock Interpolated energy densities{,} correlation indicators and lower
  bounds from approximations to the strong coupling limit of DFT.
\newblock Phys Chem Chem Phys. 2017;19:6169--6183.
\newblock Available from: \url{http://dx.doi.org/10.1039/C6CP08704C}.

\bibitem{GorLev-PRB-93}
G\"{o}rling A, Levy M.
\newblock Phys Rev B. 1993;47:13105.

\bibitem{MalMirGieWagGor-PCCP-14}
Malet F, Mirtschink A, Giesbertz KJH, Wagner LO, Gori-Giorgi P.
\newblock Exchange-correlation functionals from the strong interaction limit of
  DFT: applications to model chemical systems.
\newblock Phys Chem Chem Phys. 2014 4;16(28):14551--14558.

\bibitem{KooGor-TCA-18}
Kooi DP, Gori-Giorgi P.
\newblock Local and global interpolations along the adiabatic connection of
  DFT: a study at different correlation regimes.
\newblock Theoretical chemistry accounts. 2018;137(12):166.

\bibitem{VucGorDelFab-JPCL-18}
Vuckovic S, Gori-Giorgi P, Della~Sala F, Fabiano E.
\newblock Restoring size consistency of approximate functionals constructed
  from the adiabatic connection.
\newblock J Phys Chem Lett. 2018 5;9(11):3137--3142.

\bibitem{FabGorSeiDel-JCTC-16}
Fabiano E, Gori-Giorgi P, Seidl M, Della~Sala F.
\newblock Interaction-Strength Interpolation Method for Main-Group Chemistry:
  Benchmarking, Limitations, and Perspectives.
\newblock J Chem Theory Comput. 2016;12(10):4885--4896.

\bibitem{GiaGorDelFab-JCP-18}
Giarrusso S, Gori-Giorgi P, Della~Sala F, Fabiano E.
\newblock Assessment of interaction-strength interpolation formulas for gold
  and silver clusters.
\newblock J Chem Phys. 2018 4;148(13):134106.

\bibitem{SeiGiaVucFabGor-JCP-18}
Seidl M, Giarrusso S, Vuckovic S, Fabiano E, Gori-Giorgi P.
\newblock Communication: Strong-interaction limit of an adiabatic connection in
  Hartree-Fock theory.
\newblock The Journal of Chemical Physics. 2018 Dec;149(24):241101.
\newblock Available from: \url{https://doi.org/10.1063/1.5078565}.

\bibitem{DaaGroVucMusKooSeiGieGor-JCP-20}
Daas TJ, Grossi J, Vuckovic S, Musslimani ZH, Kooi DP, Seidl M, et~al.
\newblock Large coupling-strength expansion of the M{\o}ller--Plesset adiabatic
  connection: From paradigmatic cases to variational expressions for the
  leading terms.
\newblock The Journal of chemical physics. 2020;153(21):214112.

\bibitem{GorSav-JPCS-08}
Gori-Giorgi P, Savin A.
\newblock J Phys: Conf Ser. 2008;{117}:012017.

\bibitem{Sav-CP-09}
Savin A.
\newblock Chem Phys. 2009;{356}:91.

\bibitem{VJCTC19}
Vuckovic S.
\newblock Density functionals from the multiple-radii approach: analysis and
  recovery of the kinetic correlation energy.
\newblock Journal of chemical theory and computation. 2019;15(6):3580--3590.

\bibitem{VG19}
Gould T, Vuckovic S.
\newblock Range-separation and the multiple radii functional approximation
  inspired by the strongly interacting limit of density functional theory.
  2019;151(18):184101.

\bibitem{BahmannKaupp2015}
Bahmann H, Kaupp M.
\newblock {Efficient Self-Consistent Implementation of Local Hybrid
  Functionals}.
\newblock J Chem Theory Comput. 2015;11(4):1540--1548.

\bibitem{Lie-PLA-79}
Lieb EH.
\newblock Phys Lett. 1979;{70A}:444.

\bibitem{LieOxf-IJQC-81}
Lieb EH, Oxford S.
\newblock Int J Quantum Chem. 1981;{19}:427.

\bibitem{Per-INC-91}
Perdew JP.
\newblock In: Ziesche P, Eschrig H, editors. Electronic Structure of Solids
  '91. Berlin: Akademie Verlag; 1991. .

\bibitem{LewLieSei-PRB-19}
Lewin M, Lieb EH, Seiringer R.
\newblock Floating Wigner crystal with no boundary charge fluctuations.
\newblock Physical Review B. 2019;100(3):035127.

\bibitem{CotPet-arxiv-17}
Cotar C, Petrache M.
\newblock Equality of the jellium and uniform electron gas next-order
  asymptotic terms for Coulomb and Riesz potentials.
\newblock arXiv preprint arXiv:170707664. 2017.

\bibitem{LLS22}
Lewin M, Lieb EH, Seiringer R.
\newblock Improved Lieb-Oxford bound on the indirect and exchange energies.
\newblock arXiv preprint arXiv:220312473. 2022.

\bibitem{RasSeiGor-PRB-11}
R\"as\"anen E, Seidl M, Gori-Giorgi P.
\newblock Phys Rev B. 2011;{83}:195111.

\bibitem{SBKG22}
Seidl M, Benyahia T, Kooi DP, Gori-Giorgi P.
\newblock The Lieb-Oxford bound and the optimal transport limit of DFT.
\newblock arXiv preprint arXiv:220210800. 2022.

\bibitem{VucFabGorBur-JCTC-20}
Vuckovic S, Fabiano E, Gori-Giorgi P, Burke K.
\newblock MAP: an MP2 accuracy predictor for weak interactions from adiabatic
  connection theory.
\newblock Journal of chemical theory and computation. 2020;16(7):4141--4149.

\end{thebibliography}

\end{document}